\documentclass[conference]{IEEEtran}
\usepackage{cite}
\usepackage{graphicx}
\usepackage{algorithmic}
\usepackage{algorithm}
\usepackage{subfigure}
\begin{document}

\title{Self-organization of nodes using bio-inspired techniques for achieving small world properties}

\author{\IEEEauthorblockN{Rachit Agarwal\IEEEauthorrefmark{1}, Abhik Banerjee\IEEEauthorrefmark{1}\IEEEauthorrefmark{2}, Vincent Gauthier\IEEEauthorrefmark{1}, Monique Becker\IEEEauthorrefmark{1}, Chai Kiat Yeo\IEEEauthorrefmark{2} and Bu Sung Lee\IEEEauthorrefmark{2}}
\IEEEauthorblockA{\IEEEauthorrefmark{1}Lab. CNRS SAMOVAR UMR 5157, Telecom Sud Paris, Evry, France}
\IEEEauthorblockA{Emails: {\{rachit.agarwal, vincent.gauthier, monique.becker\}}@telecom-sudparis.eu}
\IEEEauthorblockA{\IEEEauthorrefmark{2} CeMNet, School of Computer Engineering, Nanyang Technological University, Singapore}
\IEEEauthorblockA{Emails: \{abhi0018, asckyeo, ebslee\}@ntu.edu.sg}}

\maketitle

\begin{abstract}

In an autonomous wireless sensor network, self-organization of the nodes is essential to achieve network wide characteristics. We believe that connectivity in wireless autonomous networks can be increased and overall average path length can be reduced by using beamforming and bio-inspired algorithms. Recent works on the use of beamforming in wireless networks mostly assume the knowledge of the network in aggregation to either heterogeneous or hybrid deployment. We propose that without the global knowledge or the introduction of any special feature, the average path length can be reduced with the help of inspirations from the nature and simple interactions between neighboring nodes. Our algorithm also reduces the number of disconnected components within the network. Our results show that reduction in the average path length and the number of disconnected components can be achieved using very simple local rules and without the full network knowledge.
\end{abstract}


\IEEEpeerreviewmaketitle

\section{Introduction}\label{sec:intro}
Decades of research and vast implementation of wireless networks \cite{Akyildiz}, has led it to grow tremendously, thereby creating performance issues and the need for manageability and scalability. Due to manageability, scaling and the need of better performance of network, it is important that the node is autonomous. Researches have proved, autonomous behavior not only helps in scalability and manageability but also helps in achieving global consensus using local information, cost efficient topology deployment and maintenance and the evolution of the network over time \cite{Dressler}. Due to the autonomous behavior of the nodes these models were mostly decentralized and inspirations from nature were drawn.

Inspired by the experimental work of \cite{Milgram}, in \cite{Watts} Watts et al proposed that the average path length (\emph{APL}) of a regular wired network can be reduced by introducing few long-range links within the network and developed the concept of small world networks. They proved that by rewiring few connections within the network the \emph{APL} can be considerably reduced while the clustering coefficient (\emph{CC}) can mostly be preserved. Several works since then were involved in achieving small world characteristics in the network and addressed scaling and performance issues \cite{Barabasi,BarabasiAlbert}. Nevertheless, rewiring of links in wireless networks is still a relatively hard task to achieve due to the spatial nature of network and distance limited property of the wireless links \cite{Helmy}. However, these shortcuts can be introduced in many ways. For example, by using directional antenna of the same power as of omnidirectional antenna or by increasing the omnidirectional transmission range of the nodes which causes early death or by adding another transmission antenna for beamforming. In \cite{Sharma}, few long wired links were introduced while in \cite{GuidoniLoureiro}, few special nodes with higher omnidirectional transmission range were used. Despite above mentioned techniques for achieving shortcuts, problems like finding beam direction, beam length and determining the new neighborhood due to change in the beam properties are always associated in wireless networks. Previous researches on beamforming antennas has been concentrated on networks with uniform distribution and high-density \cite{GuidoniLoureiro,Brust,Bettstetter,Vilzmann,VilzmannWidmer} but very few among them talk about non-uniform distribution of nodes. Most of the researches, considering that all nodes beamform \cite{Bettstetter,Vilzmann,VilzmannWidmer,Kiese,Yu,Li} address connectivity very well but do not discuss the impact on small world characteristics.

In this paper, however, we discard the possibility of adding any external infrastructure and focus ourselves on how small world characteristics can be achieved in homogenous environment using beamforming antenna models \cite{Balanis}. For our current study, we have used sector model \cite{Yu}. We have applied beamforming feature to transmission antenna only, though there are researches performed on the application of beamforming to reception antenna also \cite{VilzmannWidmer,Kiese,Yu,Li}.

In sparse wireless networks, most wireless nodes are unconnected from the network. This motivates us to investigate beamforming related issues like connectivity, \emph{APL} and \emph{CC} in a sparsely distributed wireless network with the help of algorithms inspired from nature and local information. We propose that Lateral Inhibition \cite{Lawrence,Nagpal,Afek} and Flocking \cite{Reynolds} can provide us valuable insights towards a solution to the above-mentioned problems in conjunction with centrality concept of graph theory. In our parallel work we have proposed an algorithm based on traffic flow, centrality measure calculation and beamforming to achieve small world characteristics. The algorithm, however, targets densely populated networks.

Furthermore, this paper provides a brief overview of assumptions in section \ref{sec:sec1} considered for modeling our algorithm in section \ref{sec:sec2} followed by simulation setup scenario and results in section \ref{sec:sec3}. We finally conclude our work in section \ref{sec:sec4} and provide some insights to future research directions.

\section{Our Model}\label{sec:sec1}
In-order to address some of the previously mentioned issues, we focus ourselves to homogenous and autonomous deployment of wireless network nodes. This type of deployment helps us to easily apply self-organizing features, achieve global consensus with local information, choose leader randomly, make network highly fault tolerant, easily maintain network topology, reduce deployment cost and extend to incorporate mobility. As the nodes are homogenous, they all inherit beamforming capabilities but the decision to use directional antenna is decided using very simple local rules. As stated in section \ref{sec:intro}, we have used sector model to visualize our model and have assumed all transmissions of data to be synchronous.

As we try to use local information, it is first very essential to know the information and the source of the information. We say local information is the information available with node and its one-hop neighbors. Determining the one hop neighborhood is thus an essential part for the correct operation of the algorithm. Many neighborhood discovery mechanisms have already been proposed and have been carefully analyzed \cite{Vasudevan}. Not focusing on the neighborhood discovery, we limit our focus towards increasing the connectivity in an unconnected network, reducing the \emph{APL} and maintaining the \emph{CC}. We further divide our approach into two parts:
\begin{itemize}
\item[A)]	Region formation with centroid finding: To reduce message overheads and to determine the nodes that a node should beamform toward in-order to achieve the reduced \emph{APL}, \ref{sec:sec2}.
\item[B)]	Flocking inspired Beamforming: To determine nodes that beamform, direction and width of the beam and to address connectivity, \emph{APL} and \emph{CC}, \ref{sub:s2}.
\end{itemize}

\section{Algorithm}\label{sec:sec2}
\subsection{Region formation and Centroid finding}\label{sub:s1}
Closeness Centrality \cite{Freeman,Freemanlc}, determines the most important node in the network through which information can be propagated to other nodes easily and quickly in the least number of hops. Due to the global property of the Closeness Centrality, the determination of the Closeness Centrality of the node requires the nodes to know other nodes in the region. Storing information about all nodes may consume lot of space. To overcome this problem, we create logical regions and find centroid node of the region based on local information. As suggested, creation of regions not only helps in reducing the message complexity in the network but also helps in reducing the effect on the \emph{APL} due to the failure of a node, the effect of disease spread is only limited to the region thereby making the network more manageable, tolerable to the failures and efficient \cite{BrustRibeiro}. Some algorithms in this direction were centralized where Base Station (BS) based on their energy level, position, transmission power, degree or mobility of the node, example WACA \cite{BrustAndronache}, assigned regional heads. On the contrary, other algorithms were either distributed \cite{Heinzelman} or probability based \cite{Younis}.

\begin{figure}
\centering
\includegraphics[width = 75mm]{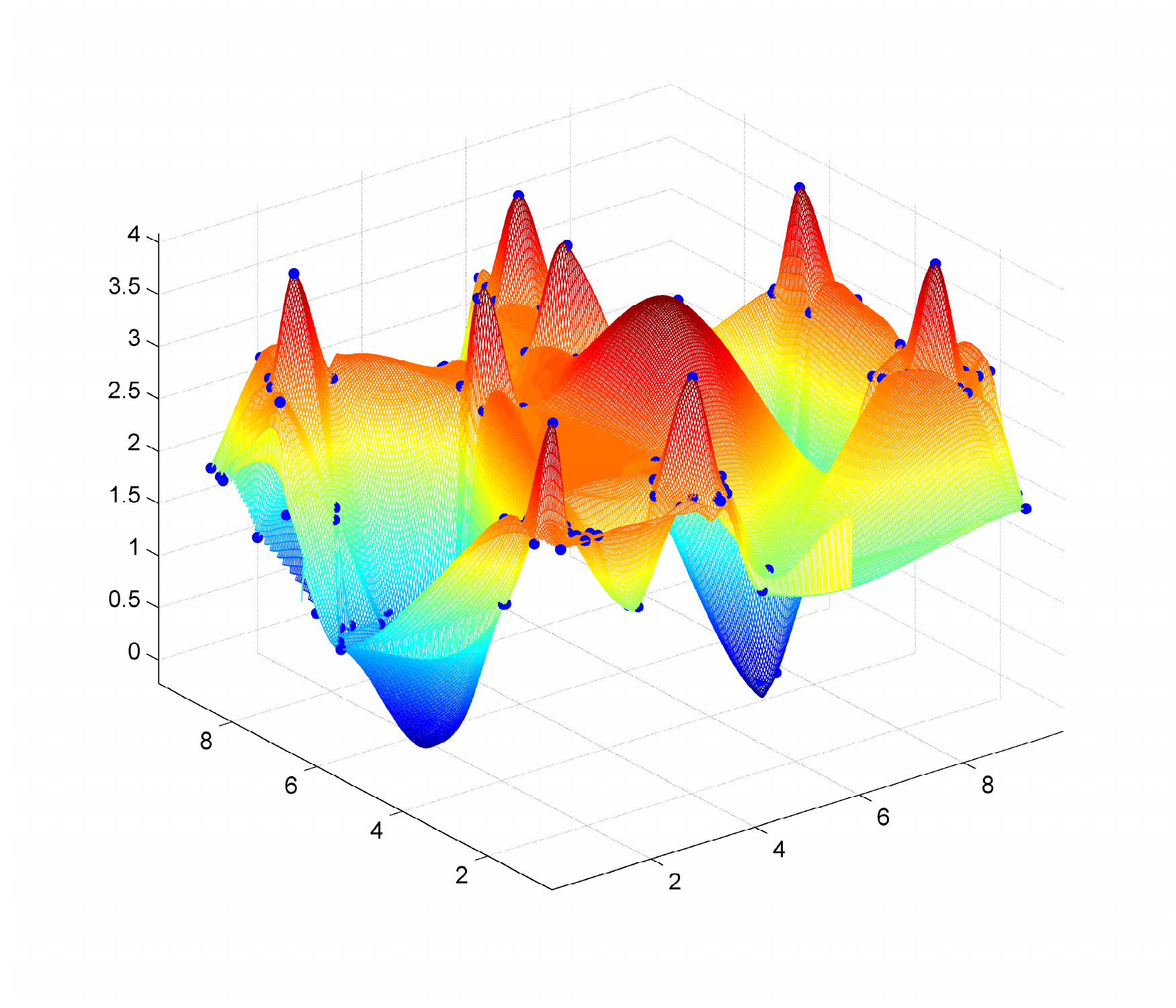}
\caption{Hopcount of the nodes. The peaks are the centroid nodes while the valleys contain nodes with the max hopcount from the centroid. Here the max hopcount=4.}
                \label{fig:gradient}
\end{figure}

As our model is distributed with nodes having only one hop information, lateral inhibition serves our purpose very well. We consider, a node broadcasts a message containing three information: the head ID to which it is associated, the hopcount from the head node and the degree of the head node it is associated with. Initially, all nodes consider themselves as head and broadcast their own information, i.e., their ID, hopcount=0 and their own degree. When the node receives information from its neighbor that has a higher degree, the node updates its leader information and broadcasts the received information instead of its own information, thereby inhibiting itself from being the regional head. When the node receives the same node degree from its neighbors, the inhibition decision of the node is based on the lower hopcount. When the hopcount is also same, then the node randomly decides for the head from the set of received information. Each node, if inhibited, increments the received hopcount by one in-order to know the exact distance from the head and virtually forms a gradient for the region (Cf. Fig. \ref{fig:gradient}). This helps in creating size-limited regions with head distributed all across the network. The node also track of the information about the distance limited heads it has received during the association phase mentioned above. This makes the node to know its head as well as have information about other head within few hops. Nodes with no neighborhood are tagged as heads because at the end of this process they remain uninhibited.

\begin{algorithm}[!ht]
\caption{Region formation and centroid finding}\label{Algo1}
\begin{algorithmic}[1]
 \FORALL {node}
	\STATE set \ensuremath{nodeStatus = uninhibited}
	\STATE set \ensuremath{virtualCoordinates = \ensuremath(x,y)}
	\STATE broadcast(\ensuremath{headID, hopCount, degree})
\ENDFOR
\REPEAT
    \STATE \ensuremath{recv}=receive(\ensuremath{headID, hopCount, nodeDegree})
    \IF {degree\ensuremath{<}nodeDegree \& hopCount\ensuremath{<}gradientSize}
        \STATE nodeStatus=inhibited \& broadcast(\ensuremath{recv})
    \ENDIF
\UNTIL {converges}
\FORALL {nodes in a region in all regions}
    \STATE \ensuremath{newCoordinate}=Centroid finding algorithm \cite{Watteyne}
\ENDFOR
\FORALL {nodes where \ensuremath{virtualCoordinates-\varepsilon<newCoordinate<virtualCoordinates+\varepsilon}}
    \STATE compute sum(\ensuremath{degree, egocentricBetweenness})
    \STATE declare the node with max sum as centroid
\ENDFOR
\end{algorithmic}
\end{algorithm}

The current technique of region formation and head selection is based on local information, but this head might not have high Closeness Centrality. Insights from \cite{Watteyne} can be used to find the centroid node within these logical regions. All the nodes in the region assign themselves randomly selected virtual coordinates. The nodes then compute the average of the virtual coordinates using the virtual coordinates of their neighborhood and broadcast it. The neighbors' intern uses these coordinates to compute new average. This process continues until all the nodes in the region have the same average. This technique reveals the coordinates of the centroid but not the ID. The nodes, in-order to identify the centroid node use their initially assigned coordinates and the newly found average coordinates. Each node then checks if the average coordinates is same as their initial virtual coordinates. If the initial virtual coordinates are within the error margin, $\varepsilon$, of the average coordinates, the node declares itself as the centroid. This might result into multiple nodes declaring themselves as the centroid. To avoid this, the decision of being centroid is also taken based on the node degree and its egocentric betweenness \cite{Everett,Daly,Marsden}. Being local measures, both degree and egocentric betweenness can be easily computed using local information. Once the centroid node is identified, the centroid information is broadcasted and the nodes update their head ID to centroid node ID and the hopcount with the hopcount to the centroid node of the region. Algo. \ref{Algo1} describes region formation and centroid identification while the Fig. \ref{fig:gradient} depicts the result of algo. \ref{Algo1} on a network shown by points in Fig. \ref{fig:gradient}.
\subsection{Beamforming}\label{sub:s2}
As discussed earlier, to achieve small world properties in wireless networks, it is essential to find the beamforming nodes, direction and the width of the beam. Flocking provides us with a valuable insight in determining the answers to these questions. Alignment, cohesion and separation rules of flocking can effectively answer these questions. In flocking, alignment rule allows a node to orient itself towards the average direction of the motion of the neighborhood, cohesion rules binds the node towards the centroid of the neighborhood while separation rule prevents the node from colliding with the neighborhood node. Similar rules inspired from flocking rules can be applied in our algorithm. For identifying beamforming nodes, we use modified alignment rule of flocking. We say that the nodes align themselves towards the decision of whether to create the beam or not. The alignment rule we apply is thus to identify peripheral nodes (\emph{P}) of the regions formed in the the previous section. The decision of being peripheral is made based on the hopcount of the neighborhood from the centroid node. If a neighbor has a hopcount less than or equal to the node's hopcount, the node declares itself as a \emph{P}. A single unconnected node is considered as a \emph{P} as it does not have any neighborhood to compare its hopcount.
\begin{figure}
\centering
\includegraphics[width = 0.4\textwidth]{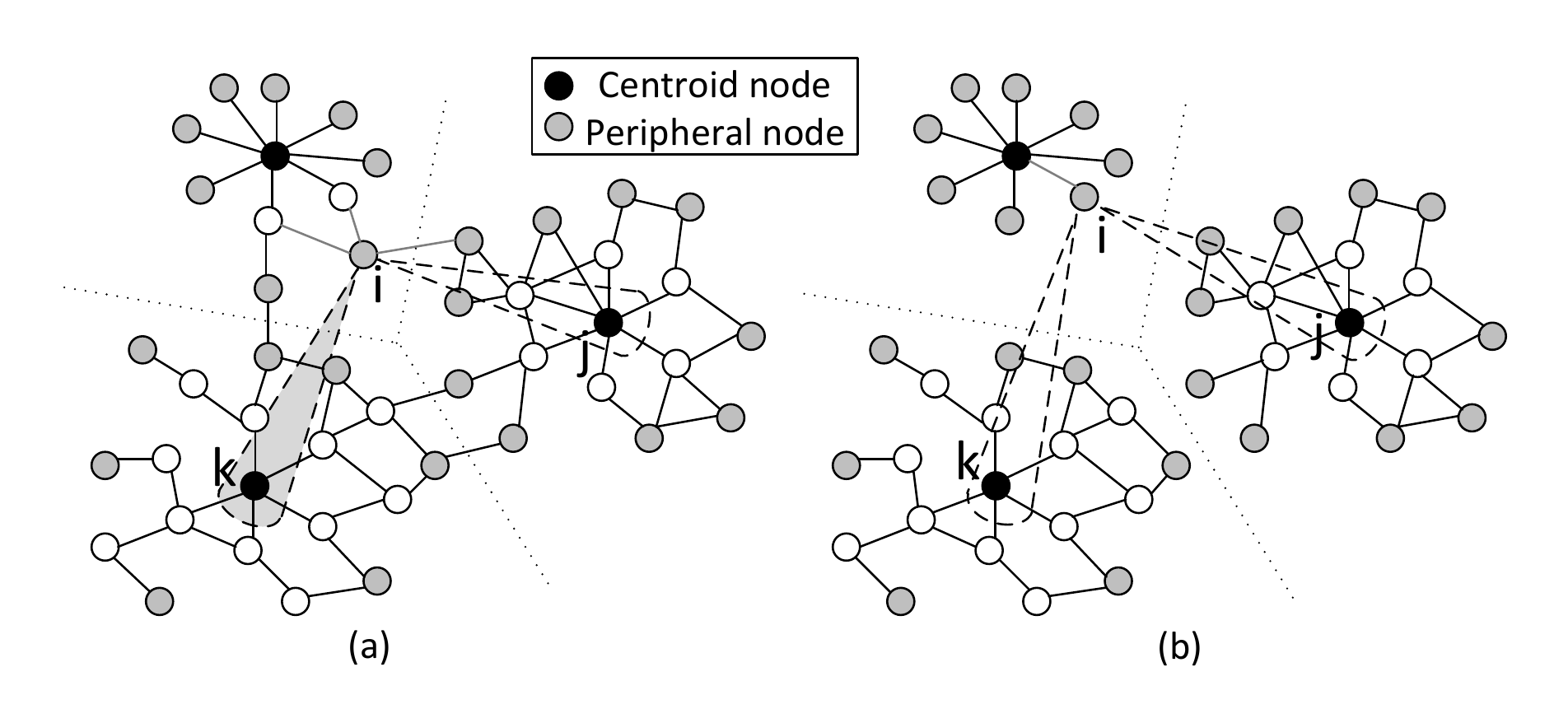}
\caption{Given gradient=3, a) One component with three regions. Node i creates beam towards centroid k instead of j because its distance to k is more, b) Three unconnected components. Node i can create beam either towards j or k as its distance to both j and k is $\infty$.}
\label{fig:f3}
\end{figure}
Once the node has been aligned, another question of choosing the direction of beam arises. To this we say, cohesion rule of flocking helps us in determining the best direction of the beam. As the nodes are homogenous, cohesion rule helps us increase the connectivity of the network also. We apply modified cohesion rule and say that beams are directed towards the centroid of other region in order to increase the connectivity, (Cf. Fig. \ref{fig:f3}). If no new centroid is found, the decision of whether to connect to self-region centroid is made. This decision depends on the hopcount from the self-region centroid as creating beams towards self-region centroid is only feasible if the \emph{P} is more than one hop away from the self-region centroid.
%\begin{figure}[ht]
%\centering
%\includegraphics[width=42mm]{f2}
%\caption{Nodes beamforming towards different region's centroid. The maximum gradient value for lateral inhibition is 3.}
%\label{fig:f2}
%\end{figure}

Considering the sector model for now, each sector in the sector model is of equal width for a given length (max gain = number of elements) and thus the beam width can easily be computed. In the sector model, we further assume that each element has the same energy consumption rate. Nodes randomly chose the number of antenna elements and use the above mentioned rules to beamform. The best beam direction and the knowledge of whether a connection to the centroid node is established is still unaddressed. As the network deployment is sparse, there can be many unconnected network components. To increase the connectivity, it is thus important for \emph{P} to find these components and be connected to them. The alignment and cohesion rule does not guarantee this coverage. Flocking's separation rule provides an insight towards this problem. We say, in-order to increase connectivity, a node creates the beam in a different direction as its neighbor peripheral nodes (\emph{$N_{pn}$}). To make this decision, if a \emph{P} creates a beam towards a centroid, it informs its neighbors about the chosen direction before it actually creates the beam. The \emph{$N_{pn}$} then tries to find a centroid node in other directions. If no centroid node is found, the decision of creating a long-range beam is dropped and the \emph{P} remains omnidirectional.

We consider that the nodes determine the direction using path length and sweeping. Consider one big connected component with multiple regions as shown in fig. \ref{fig:f3}(a). Let node i in one region create beam. From fig. \ref{fig:f3}(a), it can be seen that i can create a beam either towards j or k or towards its own regional centroid. As we know that \emph{APL} is dependent on $\sum_{i\neq j}^{N} d(i,j)$, any reduction in this value will lead to a reduced \emph{APL}. In order to have reduced \emph{APL} we propose that the node connects to the farthest centroid and can compute this information using previously stored local information. In fig. \ref{fig:f3}(a), node i is 5 hops away from k while it is 4 hops away from j. Thus, in order to have a reduced \emph{APL}, node i decides to create beam towards k. In the case when the node does not have previously stored information about the centroid nodes, the node considers the hopcount to those centroid nodes as infinite and connects to them, (Cf. Fig. \ref{fig:f3}(b)). When \emph{P} creates a beam towards the centroid to which it was connected with $\infty$ hops, the problem of asymmetric link arises. Due to this asymmetric link, \emph{P} will not know whether it has connected to the centroid of other region or not. To solve this we say, when a centroid node receives information about a \emph{P} trying to connect to it, it just for one time, to acknowledge the reception, beamform back to \emph{P}. This can be easily done after determining the angle of incidence of the beam and works well for both connected and unconnected components.
\begin{figure*}
\centering
\makebox
    {
        \subfigure[Average Path Length]
            {\includegraphics[width = 75mm]{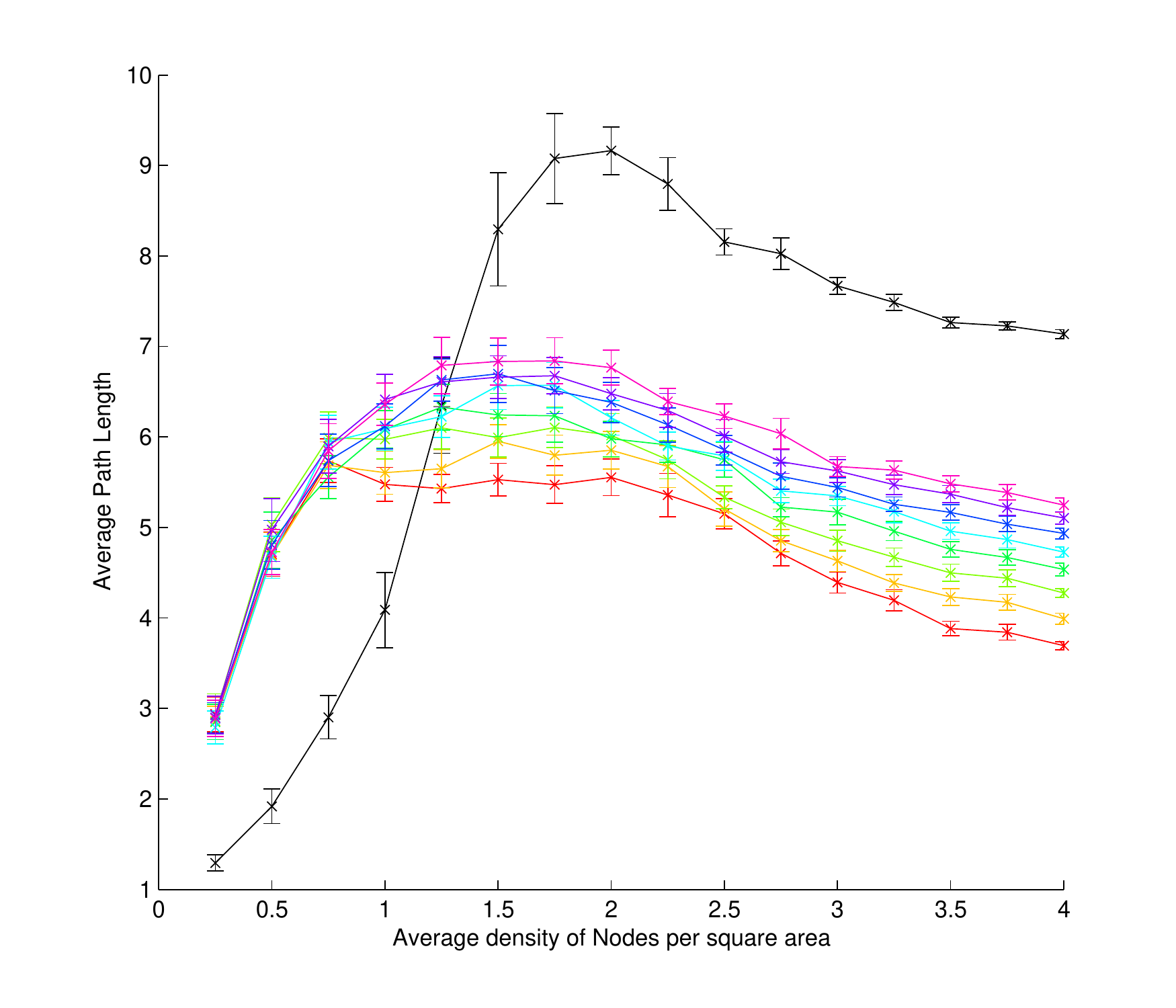}
                \label{subfig:avlsec}
            }
        \subfigure[Clustering Coefficient]
            {\includegraphics[width = 75mm]{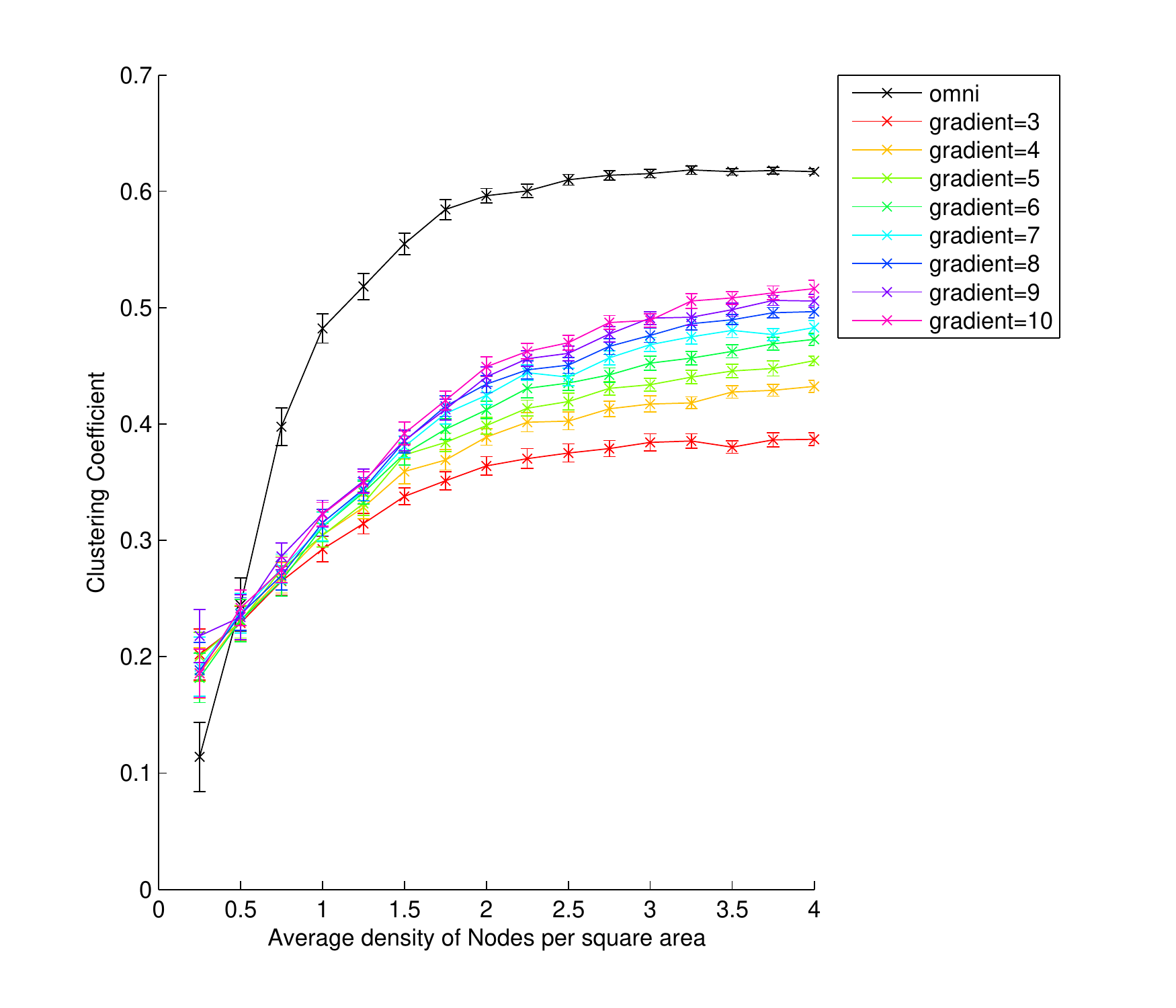}
                \label{subfig:ccsec}
            }
    }
    \makebox
    {
        \subfigure[Fraction of nodes as Peripheral]
            {\includegraphics[width = 75mm]{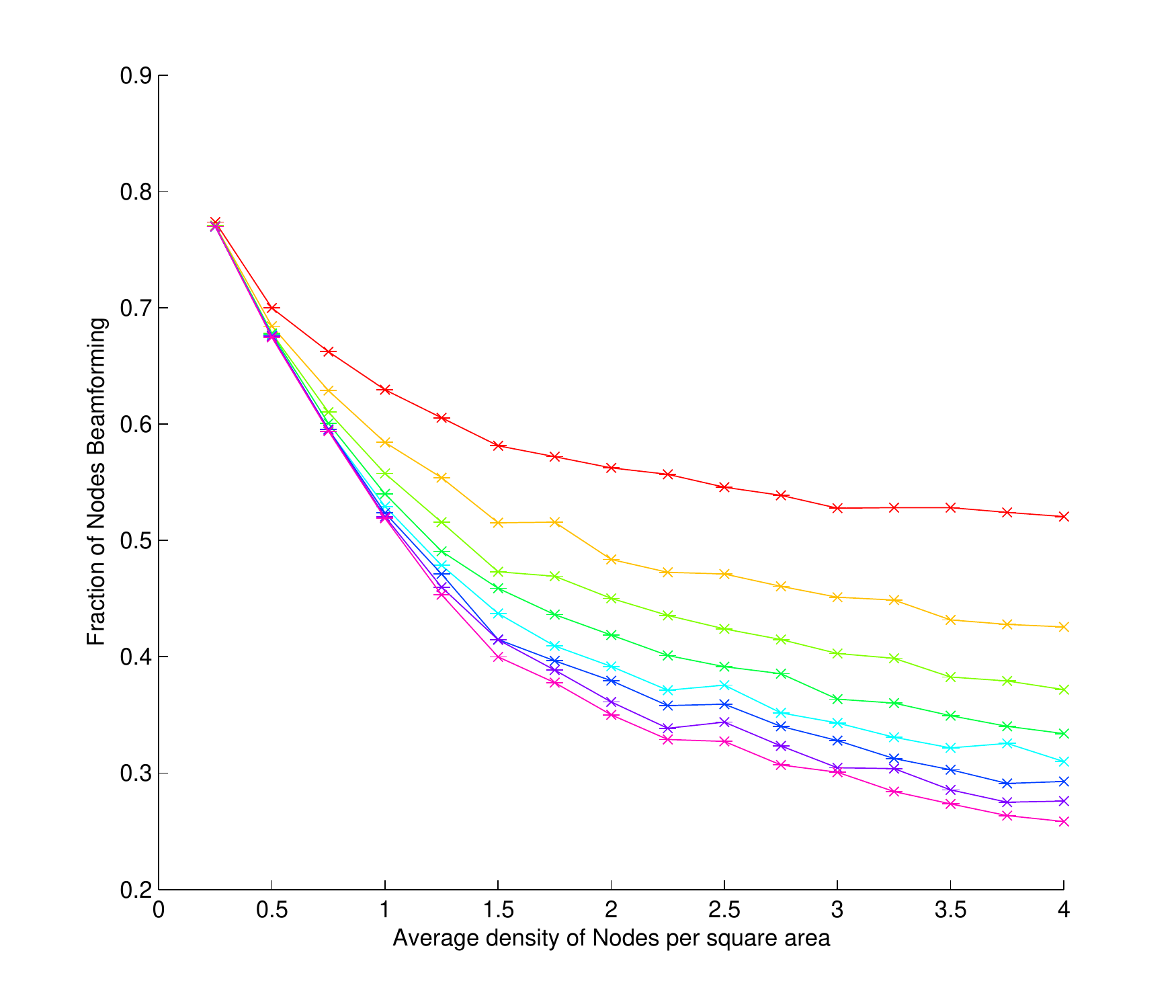}
                \label{subfig:pnodesec}
            }
       \subfigure[Fraction of nodes designated as centroid]
            {\includegraphics[width = 75mm]{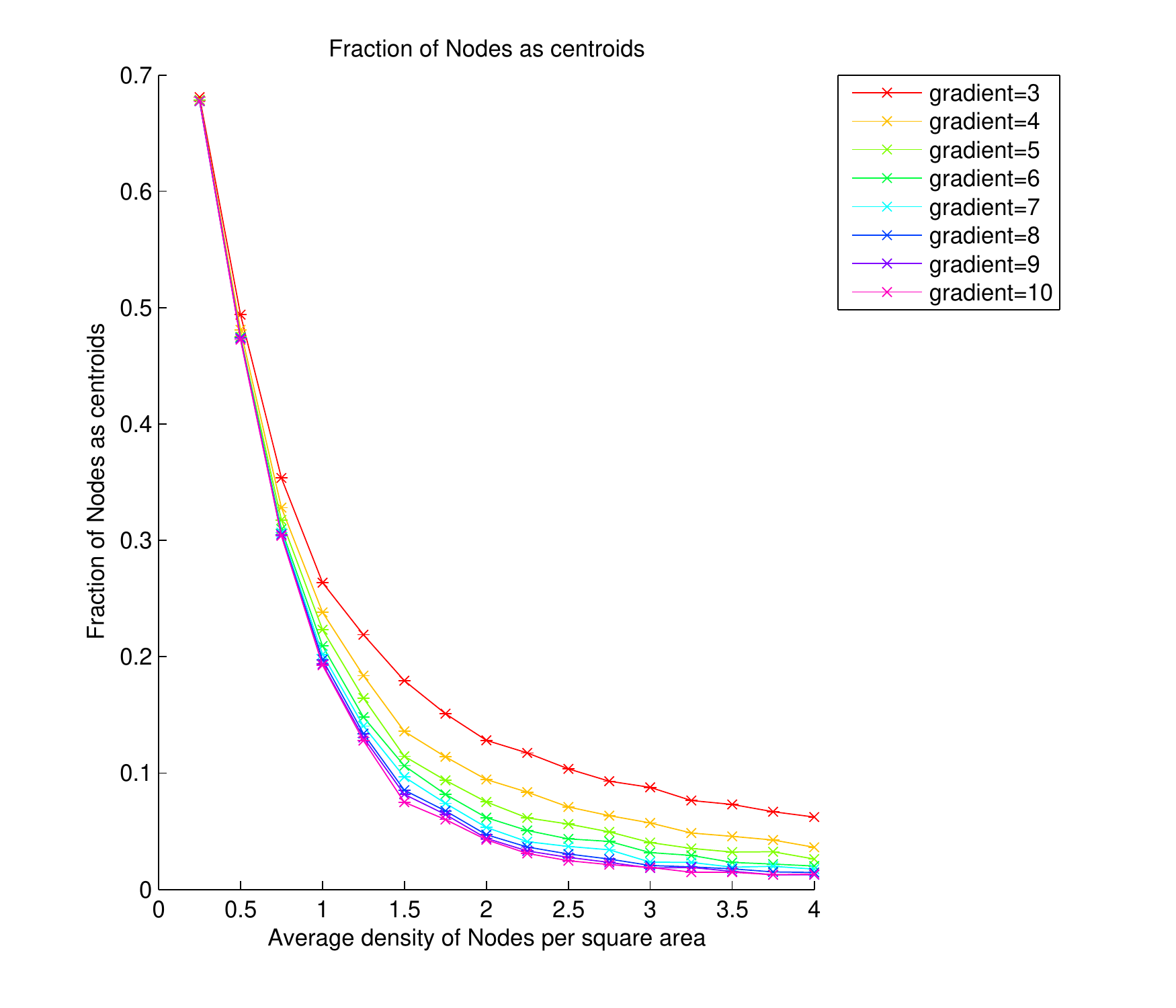}
                \label{subfig:cnodesec}
            }
    }
\caption{Results obtained for different max hopcount (gradient) for defining regions, g $\in$ [3,10], using sector model}
\label{fig:unifbiobeamsec}
\end{figure*}

\section{Simulation setup and Results}\label{sec:sec3}
In our simulation, the nodes are distributed throughout the chosen network region of 10x10. Through our simulations, we have tried to explore the effect on connectivity, average path length (\emph{APL}) and clustering coefficient (\emph{CC}) based on varying node densities and varying the gradient. Further, we have used MATLAB to simulate our model with a confidence interval of 95\%. All the results have been averaged over 50 topologies with the number of nodes varying from 20 to 400.
%    \makebox
%    {
% %\hspace{-1mm}
%        \subfigure[Hopcount of the nodes. The peaks are the centroid\newline nodes while the valleys contain nodes with the max \newline hopcount from the centroid. Here the max hopcount=4.]
%            {\includegraphics[width = 67mm]{gradient}
%                \label{subfig:gradient}
%            }\hspace{-10mm}
%        \hspace{-10mm}
%        \subfigure[Number of components in the network]
%            {\includegraphics[width = 67mm]{connectivitysec}
%                \label{subfig:connectivitysec}
%            }
%    }
%\caption{Results obtained for different max hopcount for defining regions, g $\in$ [3,10], using sector model (fig. 2a,2b, 2c, 2e, 2f)} \label{fig:unifbiobeamsec}
%\end{figure*}

We provide results obtained when sector model is used with the gradient $\in$ [3,10]. Fig. \ref{subfig:avlsec} clearly shows the effect of beamforming on \emph{APL}. \emph{APL} obtained in omnidirectional case is initially less than that obtained in the directional cases because of lower density of nodes in the component. When the directional beam is induced, due to inclusion of nodes of the other component, there is an increase in \emph{APL}. \emph{APL} for the directional case is less than that of omnidirectional case when the node density in more than 1.2 due to the fact that though nodes connect to centroid node of other regions there are some nodes that also connect to the centroid of the region in which they lie. The effect of gradient can also be seen on the peripheral nodes (\emph{P}). A low the gradient means more number of nodes are designated as \emph{P}, (Cf. Fig. \ref{subfig:pnodesec}), leading to more shortcuts and intern more reduction in the \emph{APL}. \emph{CC} however does not change much with the introduction of long-range beams, (Cf. Fig. \ref{subfig:ccsec}). For very low density networks \emph{CC} for the directional case is higher than omnidirectional case because nodes that were initially isolated now have a neighborhood. For higher density networks the effect on \emph{CC} is less for higher gradient due to less number of \emph{P}.

\begin{figure}[ht]
\centering
\includegraphics[width = 75mm]{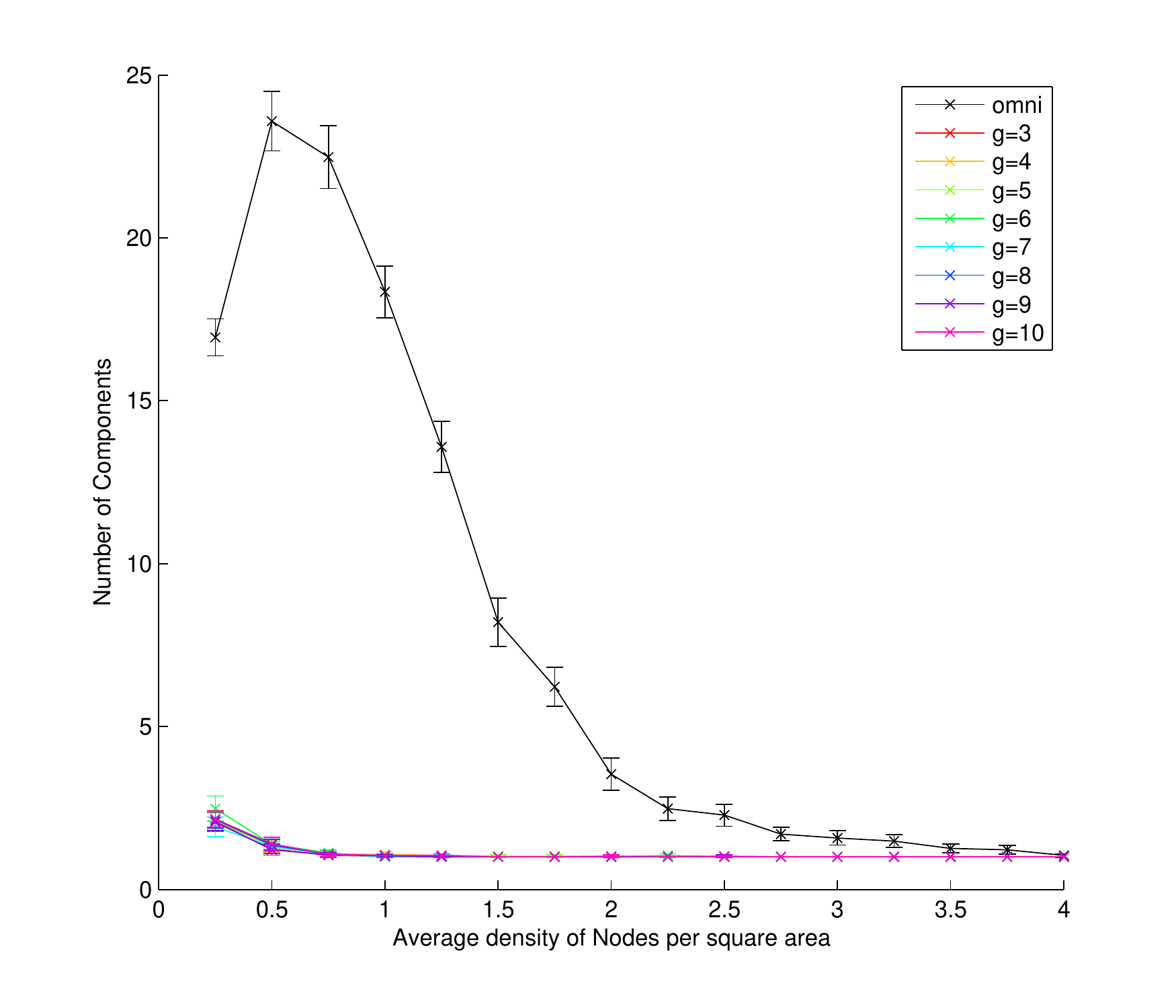}
\caption{Number of components obtained for different max hopcount (gradient) for defining regions, g $\in$ [3,10], using sector model}
\label{fig:connectivitysec}
\end{figure}
Number of components in the network can define connectivity. It can be seen for omnidirectional case from fig. \ref{fig:connectivitysec} that for very low-density networks, the number of disconnected components are more. The number of disconnected components increases to a certain maximum and then decreases as the density increases because, due to omnidirectional range being unity, nodes are more isolated in a low-density network. The connectivity is thus very low for low-density networks and there are more number of disconnected components. When the number of components decreases, the connectivity increases. For the directional case however, as nodes beamform to different components with the objective of increasing connectivity, the number of disconnected components are less than that of the omnidirectional case. The number of centroid nodes on the other hand clearly depends on the size of the gradient, (Cf. Fig. \ref{subfig:cnodesec}). For low-density network, the gradient does not matter. While as the \emph{APL} increases, the effect of gradient can be clearly seen on the number of regions. As the gradient increases, more nodes are inhibited. For a low gradient, as the density increases the effect on the number of regions is almost negligible because of increased \emph{CC} between the nodes.

The effect of gradient on the number of \emph{P} can also be seen, (Cf. Fig. \ref{subfig:pnodesec}). For low \emph{L} and low gradient, as there are more regions, more nodes become \emph{P} as they have smaller neighborhood for determination of their alignment. However, when \emph{L} and gradient are more, \emph{P} is less because there are more nodes in the region and the nodes have relatively more neighborhood to check before becoming \emph{P}. The number of \emph{P} is also related to the number of unidirectional paths and has an adverse effect on \emph{CC}.
\section{Conclusion}\label{sec:sec4}
In this paper, we have presented an algorithm for achieving small world characteristic using beamforming and bio inspired techniques in a wireless network. Our algorithm works using local information and does not require the knowledge of network. However, number of extensions to our algorithm can be visualized. The optimal gradient to choose for the determination of minimal peripheral set of nodes is clearly an extension. As we are dealing with sparse network, comparisons with different types of non-uniform distributions like Thomas point process, Mat\'{e}rn hard-core process, potential field deployment algorithm etc. can be clearly visualized as an extension. We are currently working on our algorithm extensions to mobility with asynchronous operation.
\bibliography{biblio}
\bibliographystyle{ieeetr}

\end{document}